\documentclass[12pt]{iopart}
\usepackage{graphics}      
\usepackage{graphicx}      
\usepackage{longtable}     
\usepackage{url}           
\usepackage{bm,color}            
\usepackage{ enumerate, theorem, epsfig,subfigure,setspace}

\newcommand{\nb}[1]{\textcolor{black}{#1}}
\begin{document}

\title{A Polariton Graph Simulator}

\author{ Pavlos G. Lagoudakis$^{1,2}$ and Natalia G. Berloff$^{1,3,*}$}

\address{$^1$Skolkovo Institute of Science and Technology Novaya St., 100, Skolkovo 143025, Russian Federation\\
$^2$Department of Physics and  Astronomy, University of Southampton, Southampton, SO17 1BJ, United Kingdom\\
$^3$Department of Applied Mathematics and Theoretical Physics, University of Cambridge, Cambridge CB3 0WA, United Kingdom\\
}
\ead{N.G.Berloff@damtp.cam.ac.uk}
\vspace{10pt}
\begin{indented}
\item[]7 May 2017
\end{indented}

\begin{abstract} We discuss polariton graphs as a new platform for simulating  the classical XY and  Kuramoto models.   Polariton condensates can be imprinted into any two-dimensional graph  by spatial modulation of the pumping laser.  Polariton simulators have the potential to reach the global minimum of the XY Hamiltonian in a bottom-up approach by gradually increasing excitation density to threshold or to study large scale synchronisation phenomena and dynamical phase transitions when operating above  the threshold. We consider the modelling of  polariton graphs using  the complex Ginzburg-Landau model  and derive analytical solutions for a single condensate, the XY model, two-mode model and the Kuramoto model establishing the relationships between them.

\end{abstract}

%
%
%
%
%

\section{Introduction}
Engineering a physical system to reproduce a many-body Hamiltonian has been at the heart of Richard P. Feynman's idea  of  an analogue Hamiltonian simulator \cite{feynmann}. Such simulators could address  problems that cannot be solved by a conventional Turing classical computer, which would allow us to test various models of lattice systems or discover new states of matter. Analogue Hamiltonian simulations led to the observation of  a superfluid-insulator phase transition in ultracold Bose gases  that is closely related to  metal-insulator transition in condensed-matter materials \cite{bloch}. In the last decade various other systems have been proposed as classical or quantum simulators: ultracold bosonic  and fermionic atoms  and molecular gases in optical lattices \cite{reviewUltracold,saffman,simon11,fermionic},    photons \cite{northup}, trapped ions \cite{kim10,lanyon}, superconducting q-bits \cite{corcoles},  network of optical parametric oscillators (OPOs) \cite{yamamoto11, yamamoto14}, and coupled lasers \cite{coupledlaser} among other systems.

\begin{figure}[ht]
	\includegraphics[scale=0.45]{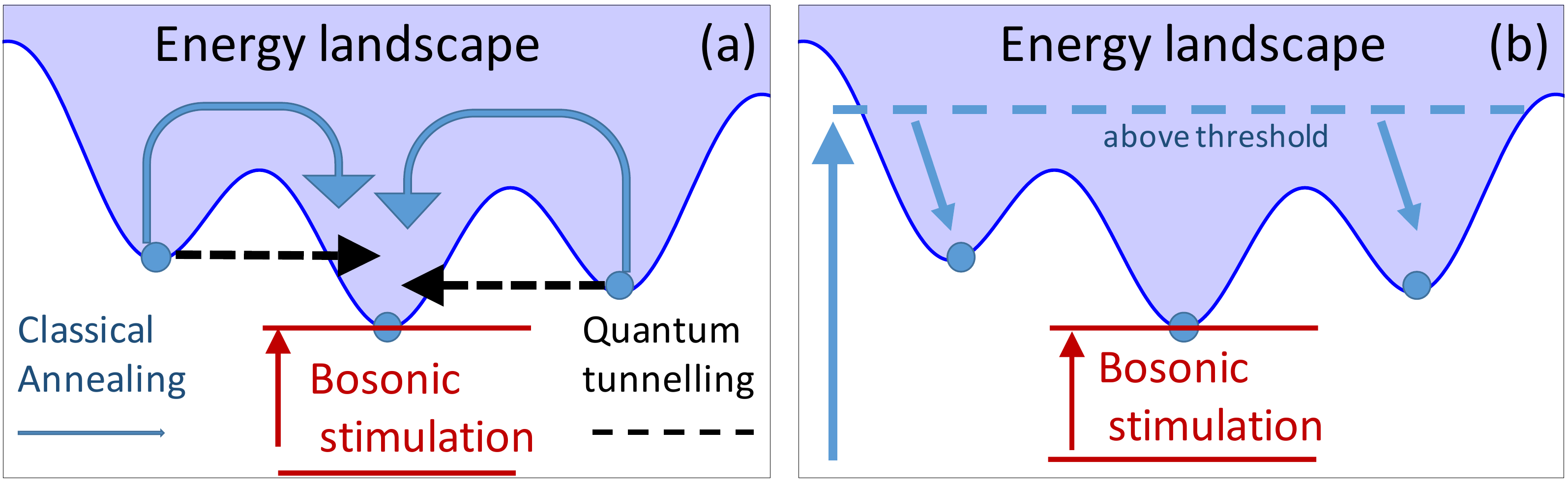}
	\centering
	\caption{(a) Schematics of different  types of annealing for finding the global minimum of the energy landscape of the simulated \nb {XY} Hamiltonian: classical annealing via the thermal activation, quantum annealing via quantum tunnelling during the adiabatic adjustment of the simulated Hamiltonian and the bosonic stimulation that leads to the system condensation at the threshold. (b) Schematics of the comparison between photon laser and polariton graph operation.  }
	\label{energy}
\end{figure}

The design of an analogue Hamiltonian simulator consists of several important ingredients \cite{dalibarOS}: (i) mapping  of the Hamiltonian of the system to be simulated into the  elements of the simulator and the interactions between them; (ii) preparation of the simulator  in a state that is relevant to the physical problem of interest: one could be interested in finding the ground or excited equilibrium state at a finite temperature; (iii) performing measurements on the simulator with the required precision. One of the platforms that have recently been explored is based on exciton-polariton lattices.  Exciton-polaritons (or  polaritons) are the composed light-matter bosonic quasi-particles  formed in the strong exciton-photon coupling regime in semiconductor microcavities \cite{weisbuch}. Under non-resonant optical excitation, free carriers relax, scatter, emit phonons and when the particle density reaches quantum degeneracy threshold, polaritons condense in the same state \cite{Kasprzak},  driven by bosonic stimulation \cite{bosonicsti}. The steady state of such a condensate is characterized by the balance of pumping and dissipation of photons that decay through the Bragg reflectors carrying all information of the corresponding polariton state wavefunction such as the energy, momentum, density,  phase and spin. This information  allows  for the in-situ characterisation of a polariton condensate in its steady state and during any dynamical transition. The non-equilibrium nature of polariton condensates gives rise to pattern formation  in these systems that has been a subject of many investigations, for review see Refs. \cite{revKeelingBerloff, revCarusotto}. Several methods for imprinting polariton lattices have been proposed. To introduce a photonic trap a partial \cite{partial}  or complete etching \cite{complete} can be used. A thin-metal film technique on a grown wafer has been used to  weakly modulate in-plane one-dimensional photon lattice \cite{film}. Exciton trap states have been explored by introducing a mechanical strain in a sample \cite{pin} or by applying  electric or magnetic fields   that  change the exciton energy \cite{ac84}.  Polariton condensates can be created at the vertices of   any two-dimensional graph by spatial modulation of the pumping source.  The first theoretical proposal \cite{keelingBerloffLattice}  and its experimental realization \cite{tosi12, tosi13} to imprint polariton condensates  in multi-site configurations were focused on the states, such as vortex lattices,  created by the outflowing polaritons from the condensate sites. Next question concerned the way the coherence is established between the various condensates. As the excitation intensity is increased from below  polaritons at the lattice site $i$ start to condense with the wavefunction $\psi_i=\sqrt{\rho_i({\bf r})} \exp[i S_i({\bf r})]$, characterized by the number density $\rho_i({\bf r})$ and a phase $S_i({\bf r})$ with the  relative phase-configuration that carries the highest polariton occupation due to the bosonic stimulation during  the condensate formation  \cite{ohadi16}. By controlling the pumping intensity and profile, the graph geometry and   the separation distance between the lattice sites one can control  the couplings between the sites and  realise various  phase configurations that minimize the  XY model as was shown in \cite{natmat17}. This gives rise to the use of the polariton graph as an analogue XY Hamiltonian simulator. The search for the global minimum of the XY Hamiltonian is via a bottom-up approach which has an advantage  over classical or quantum annealing techniques, where the global ground state is reached through either a transition over metastable excited states or via tunnelling between the states in time that depends on the size of the system. Figure \ref{energy}(a) shows the schematics of the classical thermal annealing, quantum annealing via tunnelling between the metastable states and the bosonic stimulation.

The XY model has been previously simulated by other physical systems: ultra cold atomic optical lattices \cite{struck11} and coupled photon lasers network \cite{coupledlaser}. Polariton graphs are as scalable as these platforms regarding the number of nodes it can involve. In Ref. \cite{natmat17} we have shown the graphs that consist of 45 nodes. Polariton graphs enjoy higher flexibility in engineering any geometrical configuration of nodes, since  in optical lattices and laser networks it is harder going beyond a regular lattice configuration for the arrangement of the vertices. In a microcavity used in Ref.\cite{natmat17} there is only one longitudinal mode resonant to the exciton energy and thus the system can operate  at threshold more stably than photonic laser  which  contains a large number of  longitudinal modes within the gain bandwidth.  Also, the large number of modes competing for lasing  necessitates pumping much above a threshold for stable operation. This suggests that polariton graphs have an advantage over other systems for finding the ground state of XY models. Figure \ref{energy}(b) shows the schematical difference between operating at the threshold and well above the threshold for finding the global minimum of the energy landscape which in our case is represented by the XY Hamiltonian.

The interest in using an analogue simulator for finding  the global minimum of the XY Hamiltonian is motivated by the recent result in the theory of quantum complexity:    there exist universal Hamiltonians such that  all other classical spin models can be reproduced within such a model, and certain simple Hamiltonians such as the 2D Ising model with fields on a square grid with only nearest neighbour interactions   are already universal \cite{cubittScience16}.  This suggests that hard computational problems that can be formulated as a universal Hamiltonian can be solved by a simulator that is designed for finding the global minimum of such Hamiltonian. In particular, the $XY$ model is a quadratic constrained optimization model, which is an NP-hard problem for non-convex and sufficiently dense matrices \cite{qp,natmat17}.

A more traditional use of analogue Hamiltonian simulators has  been based on modelling electrons  as they move on a lattice generated by the periodic array of atomic cores. To elucidate such a behaviour ultra cold atomic condensates were loaded  into optical lattices such as cubic-type lattices \cite{greiner}, superlattice structures \cite{sebby, folling}, triangular \cite{becker}, hexagonal \cite{becker, tarruell} and Kagome \cite{jo} lattices. Polariton graphs can easily produce such or any other ordered or disordered lattice. It was recently shown that a linear periodic chain  of  exciton-polariton condensates   demonstrate  not only various classical regimes: ferromagnetic, antiferromagnetic and frustrated spiral phases, but also at higher pumping intensities bring about novel exotic phases  that can be associated with spin liquids \cite{kalinin17a}.  \nb{Relationship between the energy spectrum of the XY Hamiltonian and the total number of condensed polariton particles  has been established in Ref.\cite{kalinin17b} where it was shown that ``particle mass residues'' of successive polariton states (defined as the difference between the masses of the individual condensates and the total mass) that occur with increasing excitation density above condensation threshold are an accurate approximation of the XY Hamiltonian's energy spectrum. Therefore, polariton graph condensate system may represent not only the ground state but also the spectral gap of the XY model. }

Our paper is organised as follows. In   Section 2 we consider the mean-field model of polariton condensates and derive analytical solutions for a single condensate. We establish the phase mapping of a polariton graph into the XY model in Section 3. In Section 4 we derive the Kuramoto model that describes the dynamics of the phases of the polariton condensates and show its relevance to the XY model. We conclude with the discussions in Section 5.
\section{\nb{An approximate analytical solution for a single condensate}}
The mean field of polariton condensates can be modelled  \cite{Wouters, Berloff} in association with atomic lasers  by writing a driven-dissipative Gross-Pitaevskii equation (aka
the complex Ginzburg-Landau equation (cGLE))  for the condensates wavefunction  $\psi({\bf r}, t):$
\begin{equation}	
	i \hbar  \frac{\partial \psi}{\partial t} = - \frac{\hbar^2}{2m}\left(1 - i \eta_d {\cal R} \right)\nabla^2\psi + U_0 |\psi|^2 \psi+
	{\hbar g_R {\cal R}}\psi
	  +\frac{i\hbar}{2} \left( R_R {\cal R} - \gamma_C \right) \psi
	\label{Initial_GL_equation}
\end{equation}
coupled to the rate equation for the density of the hot exciton reservoir, ${\cal R}({\bf r}, t)$:

\begin{equation}
\frac{\partial {\cal R}}{\partial t} = (\gamma_R + R_R|\psi|^2) {\cal R} + P({\bf r}).
\label{Main_reservoir}
\end{equation}
In these equations  $m$ is polariton effective mass, $U_0$ and  $g_R$ are the strengths of  effective polariton-polariton  and polariton-exciton interactions, respectively, $\eta_d$ is the energy relaxation coefficient specifying the rate at which gain decreases with increasing energy, $R_R$ is the rate at which the reservoir excitons enter the condensate, $\gamma_C$ is the rates of the condensate   losses, $\gamma_R$ is the redistribution
rate of reservoir excitons between the different
energy levels,  and $P$ is the  pumping  into the reservoir. 
In the limit  $\gamma_R\gg \gamma_C$  one can replace Eq. \ref{Main_reservoir} with the stationary state for the reservoir ${\cal R}=P({\bf r})/(\gamma_R + R_R|\psi|^2)$.

To non-dimensionalize the cGLE  we use
$
        \psi  \rightarrow  \sqrt{\hbar^2  / 2m U_0 \ell_0^2} \Psi,
        {\bf r}  \rightarrow  \ell_0 {\bf r}, t \rightarrow 2m t \ell_0^2/ \hbar$, where we choose $\ell_0=1\mu$ m and introduce the notations
  $g = 2 g_R/R_R,$ $\gamma = m \gamma_C \ell_0^2/ \hbar $,  $ p=m \ell_0^2 R_R P({\bf r})/ \hbar \gamma_R,\eta = \eta_d \hbar  / mR_R \ell_0^2,$ and
$ b = R_R \hbar^2  / 2m  \ell_0^2\gamma_R U_0. $

The dimensionless form of Eqs.~(\ref{Initial_GL_equation})-(\ref{Main_reservoir}) becomes the cGLE with a saturable nonlinearity

\begin{eqnarray}
i \frac{\partial \Psi}{\partial t} &=& -(1 - i\eta n_R)\nabla^2\Psi+ |\Psi|^2 \Psi+ g n_R \Psi +i \left( n_R  -\gamma \right) \Psi,
\label{Main1} \\
n_R&=&p({\bf r})/(1 + b|\Psi|^2).
\label{Main2}
\end{eqnarray}

By taking the Taylor expansion for small $|\Psi|^2$ in the expression for the reservoir $n_R$ we arrive at the more standard cGLE
\begin{equation}
i \frac{\partial \Psi}{\partial t} = -(1 - i\eta p)\nabla^2\Psi+(1-pb) |\Psi|^2 \Psi+ g p \Psi +i \left( p-\gamma  - pb|\Psi|^2\right) \Psi,
\label{cgle}
\end{equation}
where, in the view of smallness of $\eta$  we dropped $g\eta |\Psi|^2$ term. We can compare the relative strength of nonlinearities in Eqs.~(\ref{Main1}-\ref{Main2}) and (\ref{cgle}) depending of the physical quantities that define $b.$ By taking the values the system parameters typically accepted for GaAs microcavities \cite{manni,k.lagous08, tosi12} $\hbar R_R=0.1$meV$\cdot\mu$m$^2, U_0=0.02-0.04$meV$\cdot\mu$m$^2$ we obtain $b=2-4/\gamma_Rps$.  With $ \gamma_R$ on the order of 1ps$^{-1}$ we have $b$ of the order of the real nonlinearity.

%
%
%
\nb{We consider  the asymptotics and approximations of the steady state  solutions for  a single radially symmetric Gaussian pumping profile  $p(r)=p_0 \exp(-\sigma r^2)$, where $p_0$ is the maximum pumping intensity  and $\sigma$ characterises the inverse width of the Gaussian.    The Madelung transformation $\Psi=\sqrt{\rho}\exp[i S]$ relates the wavefunction to   density $\rho=|\Psi|^2$ and  velocity ${\bf u}=\nabla S$. Separating the real and imaginary parts of Eqs. (\ref{Main1}-\ref{Main2}) we arrive at the mass continuity  and the Bernoulli equations:
\begin{eqnarray}
\frac{1}{r \rho} \frac{d(r \rho u)}{dr} =  \frac{p(r)}{1+b\rho} \biggl( 1 + \frac{ \eta (r (\sqrt{\rho})')'}{r \sqrt{\rho}} -  \eta u^2  \biggr) - \gamma,\label{Mass}
\end{eqnarray}
\begin{equation}
\mu = - \frac{(\sqrt{\rho})''}{\sqrt{\rho}} - \frac{(\sqrt{\rho})'}{r\sqrt{\rho}}+{u}^2 +\rho 
+ \frac{p(r)}{1 + b \rho} \biggl( g - \frac{\eta}{r \rho} \frac{d(r \rho u)}{dr} \biggr).  
\label{Bernoulli}
\end{equation}
}

  Away from the pumping spot, where $p(r)=0$, the velocity $u=|{\bf u|}$ is given by the outflow wavenumber $k_c= const$ with $r \rho_r + \rho=-\gamma\rho r/k_c,$ which can be integrated to yield $\rho\sim  \exp[-r\gamma/k_c]/r.$ From Eq.~(\ref{Bernoulli}) at infinity we obtain
$
	\mu=k_c^2-\gamma^2/4k_c^2.
$
In the view of their  asymptotic behaviour the condensate density \nb{and velocity} can be approximated by
  \begin{equation}
  \rho(r)=\frac{a_0}{\gamma r\exp(\gamma rk_c^{-1})k_c^{-1} + \xi  -\gamma r/k_c + a_3 r^3},
  \label{rho}
  \end{equation}
  and
  \begin{equation}
 \nb{ u(r)=k_c \tanh(l r/k_c),}
  \label{u}
  \end{equation}
  utilizing their behaviour at the origin and infinity and \nb{introducing parameters $\xi, a_0, a_3$,  and $l$ that define the parametric family of  solutions. Their values should be found from the governing equations via matching asymptotics, as shown below.}
  
  \nb{We neglect $\eta$ in the view of its smallness and substitute  the expressions for the density, velocity and the pumping profile into Eqs. (\ref{Mass}-\ref{Bernoulli}). By expanding the resulting expressions about $r=0$ and  setting the term to the order  ${\cal O}(r^2)$  to zero we obtain the equations that define the unknown parameters $\xi, a_0, a_3, l$ and $k_c$ in terms of the system parameters $g, b, \gamma, p_0, \sigma$. The leading order expansion of Eq. (\ref{Mass}) and the first order expansion of Eq. (\ref{Bernoulli}) fix $\xi$ and $a_3$ as
  \begin{equation}
  \xi=\frac{a_0 b(2 l + \gamma)}{p_0-2 l - \gamma}, \qquad a_3= -\frac{\gamma^3}{2 k_c^2}.
  \label{xi}
  \end{equation}
The expansion  to  ${\cal O}(r^2)$ of Eq. (\ref{Mass}) determines $k_c$ as 
\begin{eqnarray}
k_c^2&=&[4a_0 b l^3p_0(2l+\gamma) + 3\gamma^2(8l^3 - 12 l^2 (p_0-\gamma) + (p_0-\gamma)^2 \gamma \\
\nonumber 
&+&2l (2 p_0^2 - 5 p_0 \gamma + 3 \gamma^2)]/(3 a_0 b p_0 \sigma(2 l + \gamma)^2)]
\label{kc}
\end{eqnarray}
Finally, the expansions to  ${\cal O}(r^2)$ of Eq. (\ref{Bernoulli})  define the remaining parameters $a_0$ and $l$ through two nonlinear equations
\begin{eqnarray}
k_c^2&=&\frac{a_0}{\xi} + \frac{\gamma^2 (\xi+8)}{4 k_c^2 \xi} + \frac{g p_0 \xi}{a_0b + \xi},\\
\label{eqn1}
l^2&=& \frac{a_0 \gamma^2}{k_c^2 \xi^2} + \frac{5 \gamma^4}{k_c^4\xi^2} - \frac{4 \gamma^4}{3 k_c^4 \xi} - \frac{a_0 b g p_0 \gamma^2}{k_c^2(a_0 b + \xi)^2} + \frac{g p_0 \sigma \xi}{a_0 b + \xi}.
\label{eqn2}
\end{eqnarray}
Equations  (\ref{rho}) and (\ref{u}) with the parameters defined by Eqs. (\ref{xi}-\ref{eqn2}) for the given system parameters $g, b, \gamma$ and the pumping parameters $p_0$ and $\sigma$ fully specify the approximate analytical solution of Eqs. (\ref{Mass}-\ref{Bernoulli}).   
Figure \ref{approx} shows the comparison of the approximate analytical solutions (solid lines) given by Eqs. (\ref{rho}-\ref{u}) and the numerical solutions (dashed lines)   of Eqs. (\ref{Mass}-\ref{Bernoulli}) for $b=1.5, \gamma=0.2, g=0.5$ and two sets of parameters specifying the pumping profile (a) $p_0=5, \sigma=0.2$ and (b) $p_0=10, \sigma=0.4$. The values specifying velocity are $k_c=1.65885, l=0.541858$ for (a) and $k_c=1.99239, l=0.909976$ for (b). Both analytical density and velocity profiles are in  an excellent agreement with the numerical solutions.
 \begin{figure}[ht]
	\includegraphics[scale=0.5]{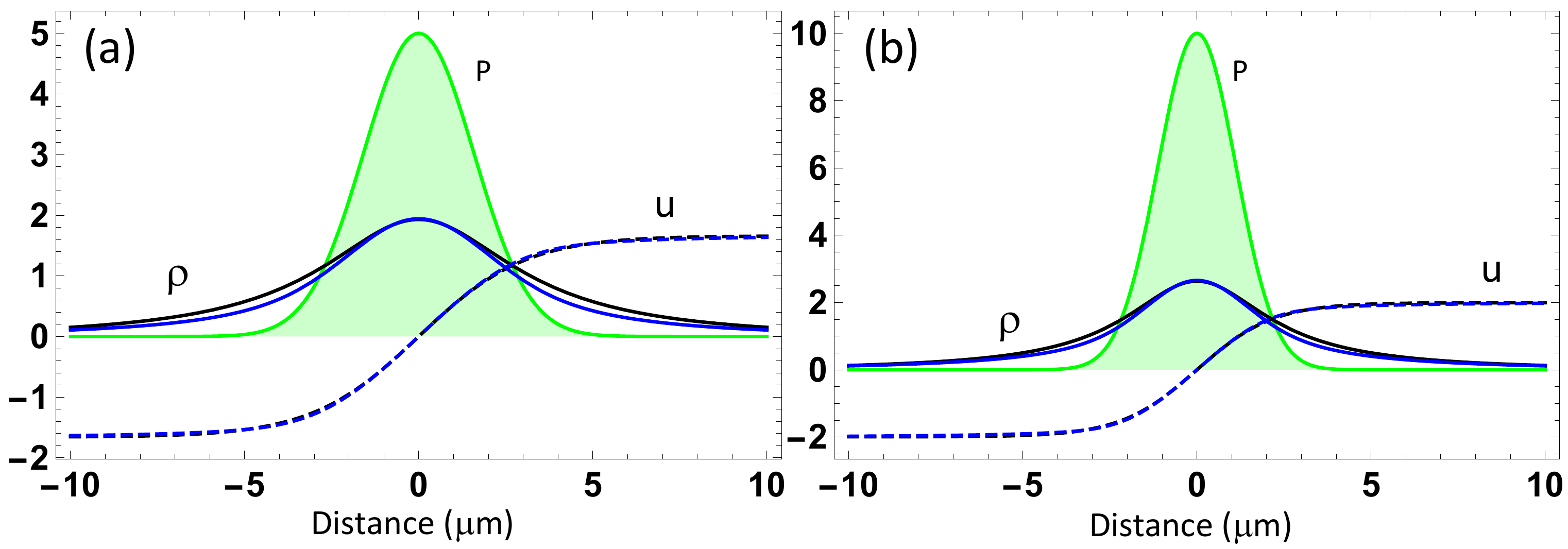}
	\centering
	\caption{Approximate analytical (blue lines) and numerical (black lines) solutions for density (solid lines) and velocity (dashed lines) of Eq. (\ref{cgle}) for the pumping profile given by $p(r)=p_0 \exp(-\sigma r^2)$ (green shaded area). The system parameters are   $b=1.5, \gamma=0.2, g=0.5$ and (a) $\sigma=0.2$, $p_0=5$; (b) $\sigma=0.4$, $p_0=10$. }
	\label{approx}
\end{figure}
}
\section{Mapping of phases into the classical XY Model}
In the previous section we obtained solutions of the governing equation Eq.~(\ref{cgle}) for a single pumping Gaussian spot. Spatial light modulator can be used to pump condensates at the vertices of a distributed graph via
\begin{equation}
p({\bf r}) = \sum_{i=1}^N p_i \exp[-\sigma_i |{\bf r}-{\bf r}_i|^2],
\label{pumpN}
\end{equation}
\nb{where $p_i$ stands for the pumping intensity at the center of the spot at position ${\bf r}={\bf r}_i$.  In what follows we shall assume that all vertices are pumped identically, so that $p_i=p_0$, $\sigma_i=\sigma$ for all $i=1,\cdot\cdot\cdot, N$.}
To the leading order and assuming that all condensates are well-separated we can approximate the resulting condensate wave function, $\psi_N$,  as
$\Psi_N({\bf r},t)\approx\sum_{i=1}^N \Psi_0(|{\bf r}-{\bf r}_i|)\exp(i \theta_i)$, where $\Psi_0=\Psi_0(r)$ is the solution of the stationary Eq. (\ref{cgle}) for a single localized radially symmetric condensate pumped by $p(r)= p_0 \exp(-\sigma r^2), $  \nb{found in the previous section
\begin{equation}
\Psi_0(r)=\sqrt{\rho_0(r)} \exp\left[i \frac{k_c^2}{l} \log \cosh \biggl( \frac{l}{k_c} r\biggr) \right],
\end{equation}
where $\rho_0(r)$ is given by Eq. (\ref{rho}).}
To find the total amount of matter ${\cal M}$ we write
\begin{eqnarray}
{\cal M}&=&\int |\Psi_N|^2\, d{\bf r} = \frac{1}{(2\pi)^2}\int |\tilde\Psi_N({\bf k})|^2 d{\bf k},\label{number}\\
\tilde\Psi_N({\bf k})&=&\int \exp(-i{\bf k}\cdot {\bf r}) \Psi_{N}({\bf r})\, d{\bf r} = \tilde\Psi_0(k)\sum_{i=1}^N \exp(i{\bf k}\cdot {\bf x_i} + i \theta_i),
\label{fourier}
\end{eqnarray}
where $\tilde{\Psi_0}(k)=2 \pi \int_0^\infty \Psi_0(r) J_0(k r) r dr$ and $J_0$ is the Bessel function. The total mass becomes
\begin{eqnarray}
{\cal M}&=&2 \pi N\int_{0}^\infty|\Psi_0|^2 r dr + \sum_{i<j} J_{ij} \cos(\theta_i-\theta_j),\label{mm}\\
J_{ij}&=&\frac{1}{\pi}\int_0^\infty|\tilde\Psi_0(k)|^2 J_0(k |{\bf r}_i-{\bf r}_j| ) k \,dk.\label{jj}
\end{eqnarray}
Since the system maximizes the total number of particles given by Eq. (\ref{mm}), this is equivalent to minimising the XY Hamiltonian functional ${\cal H}_{XY}=-\sum^n_{i<j}J_{ij}\cos\theta_{ij}$ \cite{natmat17}. The main contribution to the integral defining $\tilde{\Psi_0}(k)$ is from $k=k_c$, where $k_c$ is the outflow wavevector from the pumping site fully determined by the pumping profile \cite{ohadi16, natmat17}.

\begin{figure}[]
\centering
      \includegraphics[width=6in,angle=90]{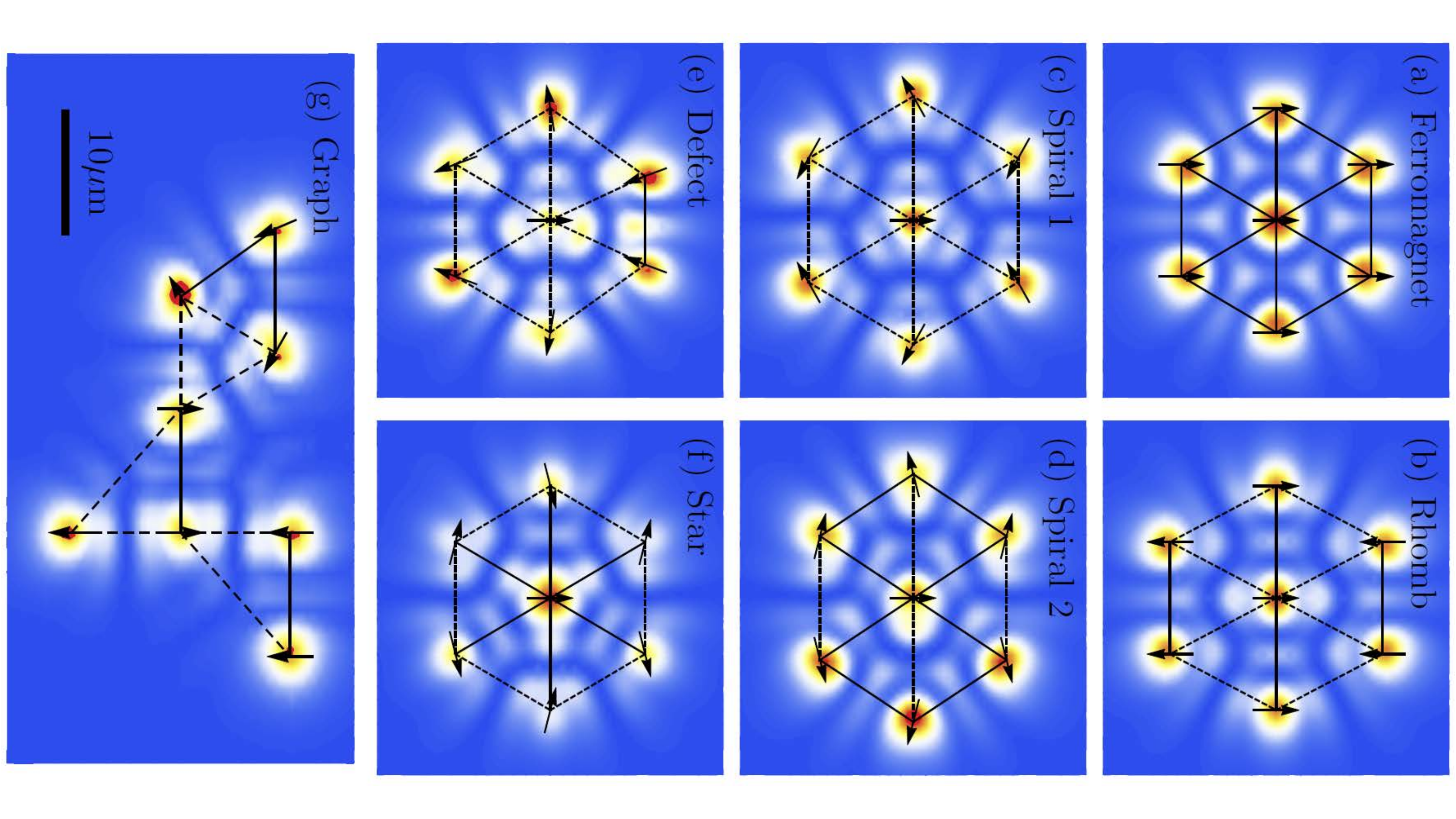}
\caption{Theoretical prediction of  classical magnetic spin configurations showing contour plots of the normalised density and spin orientations (arrows): (a) ferromagnet, with all $J_{ij}>0$, (b) rhomb, with all horizontal $J_{ij}>0$, (c) spiral 1, with all $J_{ij}<0$, (d) spiral 2, with all horizontal $J_{ij}<0$, (e) defect, with a single ferromagnetic coupling (top edge), (f) star, with $J_{ij}>0$ from central vertex to all its neighbours, (g) graph, fully disordered system. The density contour plots show $|\Psi|^2$,  with $\Psi=\sum \Psi_i$.  The individual wave functions $\Psi_i$ are analytic approximations as described in the text with $b=1.5, g=0.5, p_0=3, \sigma=0.2$.}
     \label{locking}
      \end{figure}
      
Figure \ref{locking} shows the density contour plots (normalised) and the spin orientations (arrows) representing the relative phases, with the incoherent pump spots located at the vertices. The coupling between the adjacent polariton sites can be made ferromagnetic ($J_{ij}>0$, solid lines) or antiferromagnetic ($J_{ij}<0$, dashed lines) by either varying the distance between the pumping spots, as illustrated in Figs. \ref{locking}(a-e,g) or by changing the pumping parameters of the individual spots as shown in Fig. \ref{locking}(f). Spin configuration, such as ferromagnetic, where all $J_{ij}>0$ [Fig. \ref{locking}(a)], rhombic, where all horizontal $J_{ij}>0$ [Fig. \ref{locking}(b)], spiral 1, where all $J_{ij}<0$ [Fig. \ref{locking}(c)] and spiral 2, where all horizontal $J_{ij}<0$ [Fig. \ref{locking}(d)] can be obtained by controlling independently the distance and therefore the coupling along two directions of the lattice. In atomic optical lattices, this can be achieved via an elliptical shaking of the lattice and the  spin configurations  on Fig.~\ref{locking}(a-d) were demonstrated for trapped atomic condensates \cite{struck11}.

The ultimate advantage of polariton graphs for quantum simulations is  the potential to control both the sign and the strength of any coupling, $J_{ij}$, by tuning the distance between polariton sites, or the characteristics of the pumping spots (the intensity, $p_0$, or the  inverse width of the Gaussian, $\sigma$) leading to more exotic phases. We illustrate the control over an individual $J_{ij}$ on the seven-vertex graphs of Figs. \ref{locking}(e,f). In Figure \ref{locking}(e) we  utilise control over an individual $J_{ij}$ by tuning the distance between two vertices and introduce a single ferromagnetic coupling (defect edge) into an otherwise antiferromagnetic configuration. In Figure \ref{locking}(f) we control the coupling of a single polariton site (central vertex) to all its neighbours by tuning the intensity of its pumping spot and switch them to ferromagnetic in an otherwise antiferromagnetically coupled graph (star configuration). Finally, polariton graphs allow for fully disordered systems to be addressed as shown in Fig. \ref{locking}(g).

\section{Kuramoto Model}
The cGLE can be reformulated as the  Kuramoto model  which is a  paradigm for
a spontaneous emergence of collective synchronization and that has been widely used to understand the topological organization of real complex systems from neural networks to power grids \cite{reviewKuramoto}.
In this context the polariton lattice describes collective dynamics of $N$  coupled phase oscillators with phases $\theta_i(t)$,  characterized by the natural frequencies $\omega_i$ which are associated with the chemical potential of  individual condensates.

  To show this  we adopt a two-mode model that neglects the spacial variations and represents the network of interacting polariton condensates via the radiative couplings  ${\cal J}_{ij}$ between $i$th and $j$th condensates
\begin{equation}
i\Psi_{it}=|\Psi_i|^2\Psi_i  + \biggl(\frac{(g+i)p}{1 + b|\Psi_i|^2}-i\gamma\biggr)\Psi_i + i \sum_j {\cal J}_{ij} \Psi_j.
\label{np}
\end{equation}
We neglect  the blueshift $g$ and without loss of generality let $\gamma=1$.  For the densities and phases of the individual condensates we obtain
\begin{eqnarray}
\frac{1}{2}\dot{\rho}_i(t)&=&\frac{p\rho_i}{1+ b\rho_i} -\rho_i + \sum {\cal J}_{ij} \sqrt{\rho_i\rho_j}\cos\theta_{ij}\nonumber\\
\dot{\theta}_i(t)&=&-\rho_i -\sum {\cal J}_{ij} \frac{\sqrt{\rho_j}}{\sqrt{\rho_i}} \sin\theta_{ij}.
\label{2dmodel}
\end{eqnarray}
The
radiative coupling ${\cal J}_{ij}$
 is due to the interference of the
condensates from different pumping spots \cite{rubo12}, and are such that ${\cal J}_{ij}\ll \rho_i$ for any $j$.
First, we shall assume that the density number dynamics is
faster than the phase dynamics, so that the densities acquire  the instantaneous steady state values $\rho_i=\rho=(p-1)/b$ to the leading order in ${\cal J}_{ij}.$
 In this case we get the equations of the phase dynamics represented by the  Kuramoto model:
\begin{equation}
\dot{\theta}_i(t) = -\rho -\sum {\cal J}_{ij} \sin\theta_{ij}.
\label{system}
\end{equation}
The equilibria of system (\ref{system})  are  the stationary points of the potential energy landscape
\begin{equation}
 V({\bf \theta})=\rho\sum_{i=1}^N \theta_i-\frac{1}{2}\sum_{i,j=1}^N {\cal J}_{ij} \cos \theta_{ij},
 \end{equation}
 so that the Kuramoto model (\ref{system}) describes the gradient flow to the minima of $V({\bf \theta})$ and, therefore, minimizes the XY Hamiltonian.

 Next, we will allow for the density variations and consider two spots only.
For the system with just two condensates we introduce the average density $R=(\rho_1+\rho_2)/2$ and the  half density difference $z=(\rho_1-\rho_2)/2$ for which the system reduces to three equations:
\begin{eqnarray}
\dot{\theta}_{12}&=&-2 z -2 {\cal J}_{12}\frac{R}{\sqrt{R^2-z^2}}\sin \theta_{12}, \nonumber\\
\dot{R}&=&p [(R+z) Q_+ + (R-z)Q_-]-2 R+2{\cal J}_{12}\sqrt{R^2-z^2} \cos \theta_{12},\nonumber\\
\dot{z}&=&p[(R+z) Q_+ - (R-z)Q_-]-2 z,
\label{jsystem}
\end{eqnarray}
 where we defined $Q_\pm=(1 + b (R\pm z))^{-1}$. Assuming that $z$ and ${\cal J}_{12}$ are small compared to $R$  we can expand these equations in small parameters $z$ and ${\cal J}_{12}$ and consider the steady state for $R$, which   to the leading order is $R=(p-1)/b$. Eliminating $z$ from Eq.~(\ref{jsystem}) we see that $\theta_{12}$ satisfies  the second-order differential equation
 \begin{equation}
 \ddot{\theta}_{12}+2\biggl(1-\frac{1}{p}+{\cal J}_{12} \cos\theta_{12}\biggr)\dot{\theta}_{12}=-4\biggl(1-\frac{1}{p}\biggr) {\cal J}_{12}\sin\theta_{12}.
 \label{two}
 \end{equation}
 As the pumping increases from the threshold value of $p_{\rm th}$ the oscillations between two condensates become damped with the rate proportional to $2\biggl(1-p^{-1}+{\cal J}_{ij} \cos\theta_{12}\biggr)$. The  relative phases lock to $0$ or $\pi$ difference depending on  whether ${\cal J}_{12}>0$ or ${\cal J}_{12}<0,$  respectively. This again agrees with the minimization of the XY Model, which for two pumping spots is $H_{XY}=-{\cal J}_{12}\cos \theta_{12}$ with minimum at $\theta_{12}=0$ if ${\cal J}_{12}>0$ and $\pi$ if ${\cal J}_{12}<0$.

   Despite many numerical and analytical studies of the Kuramoto model  on complex networks of different architectures there are many questions remain, in particular, on  the
dependence of synchronization on the system size, the relaxation dynamics of the model,  the  effects of time-delayed couplings and stochastic noise. In a large heterogeneous network may exist various synchronization phase transitions. These effects as well as the effect of  other correlations between intrinsic dynamical
characteristics and local topological properties could be addressed by the  polariton graph simulator.

\section{Conclusions}
In conclusion, we discussed polariton graphs as an analog platform for  minimizing the XY Hamiltonian and therefore emulating classical spin model with a potential of solving computationally hard problems. We demonstrated that the search for the global ground state of a polariton graph is equivalent to the minimisation of the XY Hamiltonian $H_{XY}=-\sum J_{ij} \cos(\theta_{ij})$. The theoretically predicted phase transitions explained the recent experiments for small and large scale polariton graphs \cite{ohadi16,tosi12,natmat17}. Polariton graphs offer the scalability of optical lattices, together with the potential to study disordered systems and to control both the sign and the strength of the coupling for each edge independently. Phase transitions in polariton graphs occur at the global ground state. With the recent advances in the field of polariton condensates, such as room temperature operation \cite{organic} and condensation under electrical pumping \cite{electrical}, polariton graph based  simulators offer unprecedented opportunities in addressing NP- complete and hard problems, topological quantum information processing and the study of exotic quantum phase transitions. Finally, we would like to emphasize that the word "quantum" could be attached to our proposal for a simulator to reflect the statistical nature of polariton condensates. The process of Bose-Einstein condensation is inherent to quantum statistics where a large fraction of bosons occupies the lowest quantum state, at which point macroscopic quantum phenomena become apparent. The use of the classical mean-field equations to describe the kinetics of the condensate does not negate the quantum statistic nature of its existence. At the same time, the proposed simulator has a quantum speed-up which is associated with the stimulated process of condensation i.e. an accelerated relaxation to the global ground quantum state.

\section{References}

\end{document}